# High-speed modulator with interleaved junctions in zero-change CMOS photonics

L. Alloatti[a),*], D. Cheian, R. J. Ram

*Massachusetts Institute of Technology, Cambridge, MA 02139, USA*

[a)]*Current address: Institute of Electromagnetic Fields (IEF), ETH Zurich, Zurich, Switzerland*

[*]*Electronic mail: luca.alloatti@gmail.com*

A microring depletion modulator is demonstrated with T-shaped lateral p-n junctions used to realize efficient modulation while maximizing the RC limited bandwidth. The device having a 3 dB bandwidth of 13 GHz has been fabricated in a standard 45 nm microelectronics CMOS process. The cavity has a linewidth of 17 GHz and an average wavelength-shift of 9 pm/V in reverse-bias conditions.

Single-chip digital circuits communicating directly using light have the potential to overcome the power and bandwidth limitations currently affecting a variety of digital systems from consumer devices to high-performance computers and data-centers.[1,2,3,4,5] Recently, we have demonstrated the first microprocessor with monolithic photonic interconnects.[5] The microprocessor was fabricated in the 45 nm 12SOI process of GlobalFoundries (formerly IBM) which is currently utilized in several high-performance computer (HPC) systems. All photonic components were realized without requiring any change to the fabrication flow and without violating the design rules of the process design kit (PDK) -a term that we named "zero-change CMOS".[6] This approach offers the advantage of leaving the transistors' yield and specifications unaltered, therefore avoiding expensive process-development and transistor qualification steps.[7,8] However, realizing high-performance photonic components with such material and process constraints has remained a challenge, and the bandwidth of the first demonstration was limited to 2.5 Gb/s.[5]

Currently, the 45 nm photonic toolbox includes waveguides with 5 dB/cm loss,[6] low-loss grating-couplers,[9] 32 GHz/0.023 A/W or 12.5 Gbps/0.55 A/W photodiodes,[8,10] and 5 Gbps modulators.[11] The achievable link bandwidth is therefore limited by the modulator.

In this work we have improved[11,5] the modulation bandwidth by an order of magnitude and simultaneously reduced the reverse-bias operating voltage. The modulator consists of a microring resonator whose resonant wavelength is modulated through plasma-dispersion in lateral pn-junctions, Fig. 1. The cavity has a radius of 5 $\mu$m and is etched into the crystalline silicon which is normally used in the 45 nm process for realizing the body of the transistors. The disk-like cavity exploits a whispering gallery mode concentrated in the outer half of the disk for effectively separating the optical field from the metal contacts as described in more details by Shainline et al..[11] The silicon layer has a thickness of less than 100 nm. After wafer processing and chip dicing, the substrate is thinned to facilitate optical confinement.[6]

High-speed operation of such lateral pn-junction modulators is challenging because of the inherent trade-offs between the various metrics for the modulator performance. Efficient modulation requires a high capacitance density but this must be balanced against the requirement for a small RC time-constant for high-speed performance. Similarly, the modulation efficiency improves with the Q of the cavity which occurs if the pn-junction is lightly doped but again there is a trade-off with the RC limited bandwidth of the device. In previous work on spoked-ring modulators, the average reverse-bias modulation efficiency was of 6 pm/V (0 V to -4 V), the operation wavelength was of 1263 nm, the FWHM was of ~26 GHz (corresponding to a loaded Q = $\lambda/\delta\lambda_{FWHM}$ = 8800), and the device bandwidth was in the 1 GHz to 3 GHz range.[11,12,13]

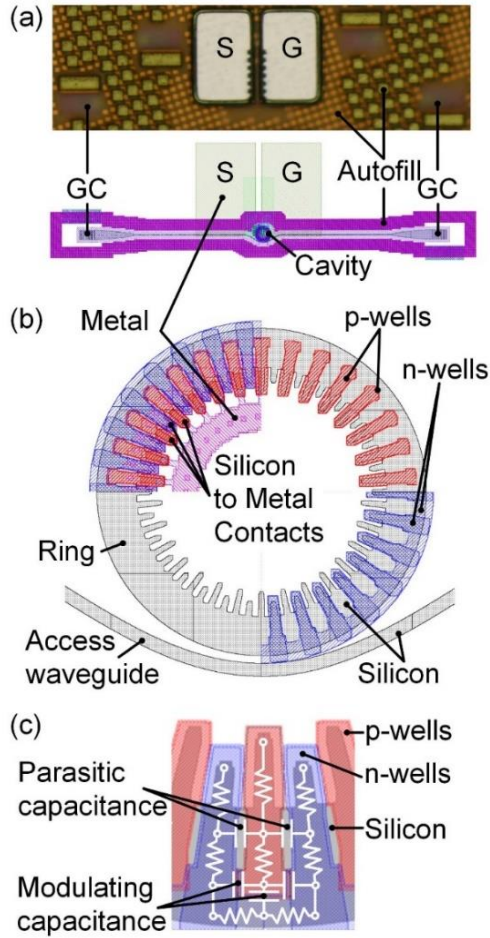

Fig. 1. Modulator geometry. (a) Microscope image of the fabricated device and full layout. Grating-couplers (GC), ground (G) and signal (S) electrodes, optical cavity and autofill structures are recognizable. (b) Zoom-in of the cavity layout. The image is divided in four quadrants with different mask layers highlighted/hidden. In the top-left quadrant most of the layers are activated including the inner metal contacts and some vias. The top-right quadrant shows the silicon and p-well layers only. The bottom-left quadrant shows the silicon layer only. The bottom-right quadrant shows the silicon and the two n-wells only. (c) Zoom-in of the well-implants showing a schematic representation of the resistances and capacitances involved. The optical mode is concentrated in the outer half of the disk. Only a fraction of the pn-junction capacitances overlaps with the optical field (modulating capacitances). The remaining capacitances are parasitic and must be minimized for decreasing the RC time constant of the circuit.

Here we leverage the lithographic precision of advanced CMOS to simultaneously improve all of these parameters. All layers have been drawn on an orthogonal (Manhattan) grid for complying with the design rules of the 12SOI process. However, the coarse grid (>200 nm) used earlier[11] for the ion-implant masks has been replaced with a fine 1 nm grid.[13] The layout was enabled by a fully-scripted photonic-design automation (PDA) tool based on Cadence with automatic DRC-cleaning.[13] The p-region is given by the superposition of two p-well implants having the shape of a T and a cumulative dose of $(9\pm2)\times10^{17}$ cm$^{-3}$. The Ts start on the inner contacts and terminate at about 300 nm from the outer radius, i.e. where the optical mode is maximum, Fig. 1(b). Two n-well implants surround the p-region forming the interdigitated junctions, Fig. 1(b). Together, their dose is smaller than $5\times10^{17}$cm$^{-3}$. The asymmetry in the p-type and n-type doping concentration helps minimize the resistance of the low-mobility, p-doped spokes while still allowing for significant depletion modulation. The spokes are T-shaped for enabling nearly abutted implants (and therefore higher capacitances) on the outer half of the disk where the mode is concentrated, while leaving a gap of about 80 nm on the inner half of the disk for diminishing the parasitic capacitance, Fig. 1(c). Source-drain (S/D) implants and silicidation complete the electrical contact to the high-frequency ground-signal (GS) electrodes.[8]

To facilitate testing, broadband grating couplers (1170 nm -1560 nm), Fig. 1(a), are used to couple TE-polarized light in and out of the waveguide and cause a loss of at about 12 dB each, although optimized grating couplers have been demonstrated elsewhere on this chip with 1.2 dB insertion loss.[9]

The current-voltage characteristic is shown in Fig. 2(a). While the dark current is smaller than 20 pA in the reverse bias range -5 V to 0 V, the modulator shows photocarrier generation when light is coupled into the active region as reported already in a number of similar devices.[14,15,16,17] Next, we characterized the tuning of the cavity wavelength when varying the reverse bias between 0 V and -4 V, Fig. 2(b) obtaining a wavelength change of 0.036 nm. To avoid the wavelength drift caused by self-heating the optical power was lowered to about -60 dBm, and the voltage was switched after recording each datapoint a few times per second. A thermally-tuned DFB laser (QDLaser model QLD1161-8030) was used. At resonance the in-fiber output power was close to the power meter noise. Using higher input powers the transmission at resonance was measured to be -11±1 dB and comparison with other devices indicated that the cavity is slightly overcoupled. The FWHM transmission is $\delta\lambda_{FWHM} = 0.079\pm0.005$ nm (17 GHz) corresponding to a loaded Q = $\lambda/\delta\lambda_{FWHM}$ = 15'000. A closed-loop circuit and a thermal heater can effectively lock the resonator to the desired wavelength as demonstrated elsewhere on the chip.[18]

The small-signal bandwidth of the device is shown in Fig. 2(c) and has been measured with a method in all similar to what described earlier;[10] in particular, the reference plane was set on the V-connector of the 50 $\mu$m pitch GS probe. For this measurement the wavelength was set to the -3 dB transmission point and the bias voltage was set to -1 V.

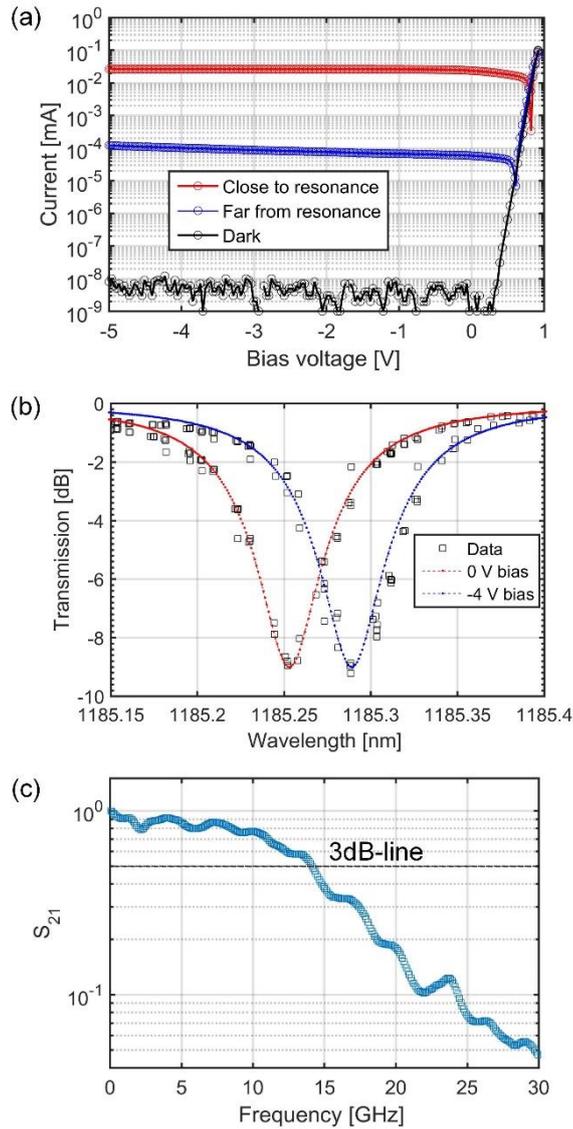

Fig. 2. Device performance. (a) Current-voltage characteristics for different illumination conditions. The dark current is smaller than 20 pA for -5 V to 0 V bias. When the modulator is illuminated it acts as a photodiode. (b) Transmission v.s. wavelength for two bias voltages (legend). The resonant wavelength changes of 0.036 nm between -4 V and 0 V. The in-waveguide incident power was reduced to about -60 dBm for diminishing wavelength detuning caused by self-heating. (c) Small-signal frequency response at -1 V bias. The 3 dB bandwidth exceeds 13 GHz.

Eye diagrams were recorded at 12.5 Gbps (PRBS length $2^{31}$-1) with an Agilent waveform analyzer (model 86108A) and a commercial 30 GHz photodiode, confirming a fast response also at large signals, Fig. 3. The insertion loss was of ~3 dB and the extinction ratio of ~5 dB with a peak-to-peak driving voltage of 2.46 V. The bandwidth was limited by the bandwidth of the pulse-pattern generator (Agilent 70843). A bias-T with 50 kHz cut off (SHF BT 65) was inserted between the photodiode and the waveform analyzer for filtering out low-frequency optical power drifts caused by mechanical vibrations of the setup. The electrical back-to-back eye with 12 dB attenuation and 50 Ω termination is shown in Fig.

3(b). The device is approximated as an open circuit and the driving voltage is calculated assuming voltage doubling at the open terminals. The in-waveguide incident optical power was between 0 dBm and 3 dBm proving that this device can sustain higher optical power levels than in our previous demonstration.[5] Along with the use of low-loss grating couplers,[9] this should enable the removal of optical amplifiers between transmitter and receiver chips in future optical links.

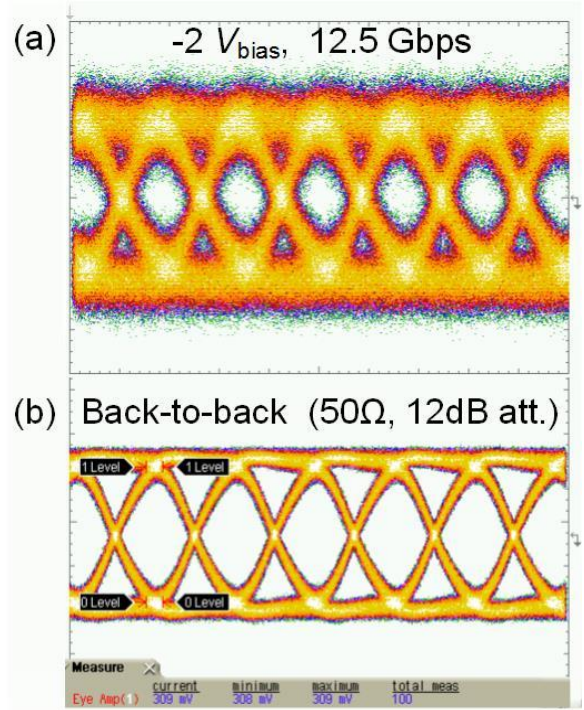

Fig. 3. 12.5 Gb/s eye diagrams. (a) Eye diagram of the modulator with a driving voltage of 2.46 $V_{pp}$ and -2 V bias. The device is approximated as an open circuit and the driving voltage is calculated assuming voltage doubling at the open terminals. The eye was recorder with a 30 GHz commercial photodiode. (b) Electrical back-to-back reference of the applied signal with 12 dB electrical attenuation and 50 Ω termination.

In conclusion, we have demonstrated a 13 GHz modulator in zero-change CMOS. Combined with the readily available high-speed photodiodes in the same 45 nm process[8] this result opens the way to monolithic chip-to-chip links with a symbol rate of 25 GBd -a tenfold improvement over previous results.[5] In the current layout we have used only four well implants out the more than ten available. Using a superposition of additional masks may therefore be used to further increase the bandwidth and the modulation depths in future iterations. For example, using the identical device geometry but optimizing the doping concentration should enable devices with similar modulation efficiency and insertion loss but with RC-limited bandwidths in excess of 20 GHz.

We acknowledge support by DARPA POEM under award HR0011-11-C-0100 and contract HR0011-11-9-0009. The views expressed are those of the authors and do not reflect the official policy or position of the DoD or the U.S. Government. We thank Amir Atabaki for performing the chip substrate transfer.

**References**

1. Weifeng Liu and Brian Vinter, Journal of Parallel and Distributed Computing **85**, 47 (2015).

2. S. Rumley, D. Nikolova, R. Hendry, Qi Li, D. Calhoun, and K. Bergman, Lightwave Technology, Journal of **33** (3), 547 (2015).

3. Solomon Assefa, William M. J. Green, Alexander Rylyakov, Clint Schow, Folkert Horst, and Yurii A. Vlasov, 2011 Optical Fiber Communication Conference (OFC), OMM6 (2011).

4. C. Gunn, VLSI Technology, 2007 IEEE Symposium on, 6 (2007).

5. Chen Sun, Mark T. Wade, Yunsup Lee, Jason S. Orcutt, Luca Alloatti, Michael S. Georgas, Andrew S. Waterman, Jeffrey M. Shainline, Rimas R. Avizienis, Sen Lin, Benjamin R. Moss, Rajesh Kumar, Fabio Pavanello, Amir H. Atabaki, Henry M. Cook, Albert J. Ou, Jonathan C. Leu, Chen Yu-Hsin, Krste Asanovic, Rajeev J. Ram, Milos Popovic, and Vladimir M. Stojanovic, Nature **528** (7583), 534 (2015).

6. Jason S. Orcutt, Benjamin Moss, Chen Sun, Jonathan Leu, Michael Georgas, Jeffrey Shainline, Eugen Zgraggen, Hanqing Li, Jie Sun, Matthew Weaver, Stevan Urosevic, Milos Popovic, Rajeev J. Ram, and Vladimir Stojanovic, Optics Express **20** (11), 12222 (2012).

7. Top500, http://www.top500.org.

8. L. Alloatti, S. A. Srinivasan, J. S. Orcutt, and R. J. Ram, Applied Physics Letters **107** (4), 41104 (2015).

9. M. T. Wade, F. Pavanello, R. Kumar, C. M. Gentry, A. Atabaki, R. Ram, V. Stojanovic, and M. A. Popovic, paper 46 presented at the Optical Interconnects Conference (OI), 2015 IEEE.

10. L. Alloatti and R. J. Ram, Applied Physics Letters **108** (7), 071105 (2016).

11. Jeffrey M. Shainline, Jason S. Orcutt, Mark T. Wade, Kareem Nammari, Benjamin Moss, Michael Georgas, Chen Sun, Rajeev J. Ram, Vladimir Stojanovic, and Milos A. Popovic, Optics Letters **38** (15), 2657 (2013).

12. M. T. Wade, J. Shainline, J. Orcutt, Sun Chen, R. Kumar, M. Georgas, R. Ram, V. Stojanovic, and M. Popovic, 2013 Conference On Optical Fiber Communication OFC (2013).

13. L. Alloatti, M. Wade, V. Stojanovic, M. Popovic, and R. J. Ram, IET Optoelectronics **9** (4), 163 (2015).

14. J. D. B. Bradley, P. E. Jessop, and A. P. Knights, Applied Physics Letters **86** (24), 241103 (2005).

15. M. W. Geis, S. J. Spector, M. E. Grein, R. T. Schulein, J. U. Yoon, D. M. Lennon, S. Deneault, F. Gan, F. X. Kaertner, and T. M. Lyszczarz, IEEE Photon. Technol. Lett. **19** (3), 152 (2007).

16. Y. Liu, C. W. Chow, W. Y. Cheung, and H. K. Tsang, IEEE Photonics Technology Letters **18** (17-20), 1882 (2006); J. K. Doylend, P. E. Jessop, and A. P. Knights, Optics Express **18** (14), 14671 (2010).

17. H. Yu, D. Korn, M. Pantouvaki, J. Van Campenhout, K. Komorowska, P. Verheyen, G. Lepage, P. Absil, D. Hillerkuss, L. Alloatti, J. Leuthold, R. Baets, and W. Bogaerts, Optics Letters **37** (22), 4681 (2012).

18. C. Sun, M. Wade, M. Georgas, S. Lin, L. Alloatti, B. Moss, R. Kumar, A. Atabaki, F. Pavanello, R. J. Ram, M. A. Popovic, and V. Stojanovic, paper C122 presented at the VLSI Circuits Digest of Technical Papers, 2015 Symposium on.